\documentclass[10pt,a4paper,english]{article}

\usepackage{latexsym}
\usepackage{epsf}
\usepackage{amssymb}
\usepackage{color}
\usepackage{amsmath} 
\usepackage[dvips]{graphicx}
\usepackage{subfig}

\makeatletter \@addtoreset{equation}{section} \makeatother

\begin{document}

\begin{center}


{\Large  {\bf Interdependent binary choices under social influence: phase diagram for homogeneous unbiased populations}\\
\vspace{.7cm}}
{\bf Ana Fern\'andez del R\'io\footnote{E-mail: {\tt ana.fernandez@invi.uned.es} \qquad  Tel: (00 34) 913987143 \qquad Fax: (00 34) 913986697  }, Elka Korutcheva\footnote{Also at G. Nadjakov Institute of Solid State Physics, Bulgarian Academy of Sciences, 1784 Sofia (Bulgaria).} and Javier de la Rubia}

\vskip 0.4truecm
\vskip 0.2cm
{\it Departamento de F\'isica Fundamental, Universidad Nacional de 
Educaci\'on a Distancia (UNED)\\
Facultad de Ciencias (UNED), Paseo Senda del Rey 9, E-28040 Madrid}
\end{center}
\vskip 2cm


\begin{center}
{\bf Abstract} 
\end{center}

\begin{quotation}
\small Coupled Ising models are studied in a discrete choice theory framework, where they can be understood to represent interdependent choice making processes for homogeneous populations under social influence. Two different coupling schemes are considered. The nonlocal or group interdependence model is used to study two interrelated groups making the same binary choice. The local or individual interdependence model represents a single group where agents make two binary choices which depend on each other. For both models, phase diagrams, and their implications in socioeconomic contexts, are described and compared in the absence of private deterministic utilities (zero opinion fields). 
\end{quotation}







\pagestyle{plain}
\section{Introduction}
\label{sec:intro}

Social scientists have been interested in the formal study
of \emph{discrete choice theory} settings for decades. These allow for the
consideration of a wide variety of problems, ranging from the study of social pathologies  \cite{Glaeser1996} to demand contexts \cite{Weisbuch2000, Weisbuch2003, Gordon2005, Michard2005, Martino2006, Gordon2008} or election results  \cite{Bernardes2002, Fortunato2007}, as well as the prevalence of different relevant social traits or opinions  \cite{ Michard2005, Galam1982, Galam2003, Nakayama2004,  Sznajd-Weron2005, Borghesi2007}. Traditionally, the effects of social influence  were neglected. Some of the work by Schelling \cite{Schelling1971, Schelling1973, Schelling1978}  and Becker \cite{Becker1981, Becker1991} are exceptions, for some of which direct links to models from condensed matter can be drawn \cite{Kirman2007, Dall’Asta2008,Gauvin2009}. Interactive setups similar to the ones that will be considered in this paper were introduced by F\"ollmer \cite{Follmer1974} and Granovetter \cite{Granovetter1978}. In the 1990s, economists such as Blume, Brock and Durlauf started developing a consistent framework in which to systematically study these problems taking into account social interactions and noted the deep connections to some models from statistical physics \cite{Blume1993, Durlauf1997, Durlauf1999, Brock2001}. These ferromagnetic setups had also been already proposed by some of the first physicist advocates of building a physics like body of knowledge for the social sciences \cite{Galam1982,Callen1974,Galam1991}. 

The traditional approach typically considers a group of agents maximising the utility function:

\begin{equation}
\label{eq:utility}
U(s_{i}, h_{i}, E_{i}(\vec{s}), \epsilon_{i}(s_{i}))= u(s_{i}, h_{i})+S(s_{i},E_{i}(\vec{s}))+ \epsilon_{i}(s_{i})
\end{equation}
 
\noindent where $u$ is the \emph{private deterministic utility}, $S$ the
\emph{social deterministic utility} and $\epsilon_{i}$ the \emph{private
random utility}. The choice $s_{i}$  made by the individual $i$ can take the values -1 or 1. Each is a component of vector $\vec{s}$, with $E_{i}(\vec{s})$ representing the $i^{th}$ agent's  subjective belief on the choices of the rest of the agents. The  \emph{idiosyncratic willingness to adopt} (hereafter IWA) or \emph{opinion field} $h_{i}$ characterises personal preferences of each agent,  its inherent predisposition towards the choice making process. Finally, $\epsilon_{i}$ is an individual and choice dependent random shock. Economic demand contexts can be studied in these setups by interpreting IWAs as a combination $b_{i}-p$, where $b_{i}$ is the \emph{idiosyncratic willingness to buy} (hereafter IWB) and $p$ the fixed price. Demand curves, i.e., the dependence on price of the \emph{fraction of adopters $\mu$} (fraction of the individuals with $s_{i}=1$), can be studied for fixed values of the rest of the parameters, as discussed, for example, in \cite{Weisbuch2000, Gordon2005, Gordon2008, Semeshenko2007}.

There is  a direct analogy that can be established between these settings (and their total utility) and condensed matter models for ferromagnetism (and their total energy). The binary choice variable $s_{i}$ can be understood as a binary spin variable and so the \emph{average choice} $s=\frac{1}{N}\sum_{i}s_{i}$ as the system's average magnetisation,  which can be easily expressed in terms of the fraction of adopters $\mu$ as $s=2\mu-1$. The action of the IWA $h_{i}$, and therefore of private deterministic utility terms, will be associated in its ferromagnetic counterpart to the action of magnetic fields. Social deterministic utility can be related to spin interactions. Random private utility is {\it translated} into the introduction of random noise. Assuming a logistic distribution for the difference of the random payoff terms $P(\epsilon_{i}(-1)-\epsilon_{i}(1)\leq z)=\frac{1}{1+e^{-\beta z}}$ identical for all agents, is equivalent to studying the ferromagnetic system using the canonical ensemble, i.e., in statistical equilibrium.  Note that in the physical model,  $\beta$ is given the statistical mechanical usual interpretation of inverse of the temperature $\beta=1/K_{B}T$, were $K_{B}$ is Boltzmann's constant. In discrete choice settings, \emph{socioeconomic temperature}, $1/\beta$ in the utility discrete choice setting, represents statistical fluctuations that can reflect an incomplete knowledge about agent's particularities, individuals' probability of behaving {\it irrationally} and/or a more fundamental uncertainty concerning \emph{free will}.  This allows for an interpretation of $\beta$ in terms of  \emph{social permeability}, which measures the potential response of individuals to social influence \cite{Galam1982}.

Under particular choices for private and social deterministic utilities, the system can be described using a mean field Ising model with constant external field \cite{Durlauf1997, Durlauf1999}, that was solved exactly in the 1930s (see for example \cite{Baxter1982}). This is only true when considering a private deterministic utility $u$ proportional to both the individual's choice and the IWA, which is considered the same for all agents (constant opinion field). Social deterministic utility will have as many terms as other agents with whom there is interaction. All agents are assumed to interact with all others and each of these terms to be proportional to the quadratic difference between the agent's opinion and its perception of the other's choice: $S=-E_{i}\left(\sum_{j\neq i}\frac{J_{ij}}{2}\left(s_{i}-s_{j}\right)^{2}\right)$. The coupling $J_{ij}$ measures how much agent $i$ wants to conform to the opinion of agent $j$ and thus the strength of social influence/magnetic interaction and is also assumed to be constant throughout the population \footnote{Dimensionless choice variables $s_{i}$ are considered, and so utility, social coupling, IWA and temperature are measured in the same units (formally equivalent to the physical scenario in natural units $\hbar=K_{B}=1$), which can be defined to suit the specific problem under consideration. They can be an attempt to quantify abstract quantities (satisfaction, wellbeing, reputation\ldots) or measure specific gains or losses (surpluses or deficits) of the individuals, for example, in money.}. Agents have \emph{rational expectations}, i.e., $E_{i}(s_{j})=s$.  Note this means all individuals will be equally compelled to align their opinions with the average choice. All agents have the same (accurate) perception of what the average behaviour of the group is. As all agents also have identical IWAs, populations are completely homogeneous. 

These are drastic simplifications concerning most socioeconomic systems of interest but are useful to understand basic qualitative features emerging from social interaction. Complexity can be added and hypothesis relaxed to make them more realistic and parallelisms to statistical physics still remain useful. It seems particularly relevant to include heterogeneity that  enables the characterisation of the group. This can be achieved using random fields and/or spin glass type models, which have been extensively studied in physics  \cite{Cohen1985, Binder1986, Mezard1987, Goldenfeld1992, Parisi1992, Young1997}. Work on these lines  in socioeconomic contexts can be found in \cite{Gordon2005, Michard2005, Gordon2008, Borghesi2007, Durlauf1997, Brock2001, Galam1991, Semeshenko2007, Galam1997, Galam2008}. Other interesting variations on the interacting scheme, such as nearest neighbours or similar \cite{Glaeser1996, Weisbuch2003, Michard2005, Bernardes2002, Fortunato2007, Sznajd-Weron2005, Follmer1974, Durlauf1997, Brock2001, Sznajd-Weron2000, Stauffer2002, Gonzalez-Avella2011}, dynamic interacting neighbourhoods \cite{Galam2008, Tessone2004} or complex networks  \cite{Bernardes2002, Gonzalez-Avella2011, Wu2004} can be used.  However, the mean field approach can be argued to be a good approximation for many problems of interest to mimic a general tendency to conform to the norm and is more tractable analytically \cite{ Gordon2005, Gordon2008, Galam1982, Nakayama2004, Borghesi2007, Durlauf1997, Durlauf1999, Brock2001, Galam1991, Semeshenko2007, Galam1997,Galam2008, Orlean1995, Nadal1998}.  Relaxing rational expectations and/or reciprocity assumptions \cite{Glaeser1996, Semeshenko2007, Orlean1995} are also interesting paths to explore, as is the study of finite size effects \cite{Galam1991, Galam1997, Galam2008}.  The study of this kind of setups as dynamical systems (which can evolve out of equilibrium) \cite{Weisbuch2003, Bernardes2002, Fortunato2007, Galam2003, Nakayama2004, Sznajd-Weron2005, Semeshenko2007,Sznajd-Weron2000, Stauffer2002, Tessone2004, Wu2004} is currently an active area of research sometimes referred to as \emph{opinion dynamics} (see \cite{Galam2008} and \cite{Castellano2009} for reviews). Similar setups can also be used to study other related problems such as resource allocation, hierarchical structures, coalition formation, social learning, and public goods games  \cite{Weisbuch2000, Martino2006, Kirman2007, Follmer1974, Blume1993, Galam2008, Gonzalez-Avella2011, Orlean1995, Nadal1998, Castellano2009, Ma2009}.

When comparing discrete choice settings subject to social influence to those without, interesting qualitatively different characteristics appear. Even the simple Ising mean field model with constant field already presents a rich statistical mechanical phenomenology \cite{Baxter1982} with interesting translations to socioeconomic language \cite{Galam1982, Nakayama2004, Durlauf1997, Durlauf1999, Brock2001,  Galam1991, Galam1997}. The qualitative average outcome (\emph{phase}) is described by its order parameter, the average choice $s$. For the \emph{unpolarised or paramagnetic phase} ($s=0$), in average half of the group is deciding for and half against. In the \emph{polarised or ferromagnetic state} ($s\neq0$), there is a greater prevalence of one of the options over the other. 

\emph{Unbiased populations} (with respect to a particular choice) are those with zero IWA for all individuals. The average outcome will be determined only by social interaction, individuals having no inherent deterministic preference for any of the choices. We can for example consider this to be the case for some \emph{fashions and traditions}, where there is no real cost or benefit for the individuals besides the social payoff provided through imitation. The unpolarised state can only be stable in this case, with a \emph{second order phase transition} for $J\beta=1$, where a system will  experience a rapid but continuous change from an unpolarised state (for low social permeability $\beta$ or high social coupling $J$) to a polarised one. In the polarised regime  ($1<J\beta$), the stable equilibrium is degenerate, as there are two identical maxima of the utility $s=\pm m$ (minima of the free energy, physically equivalent states). The previous history of the system becomes relevant and microeconomic specifications of the model may not uniquely determine its macroeconomic properties. 

When we {\it turn on} the opinion fields, these critical values of the parameters for which a second order phase transition takes place for the unbiased case, still separate regions where social utility is relevant (spontaneous magnetisation exists in the physical analog) and regions where it is not. Systems in a previously unpolarised state at $h=0$ will turn into well behaved, single stable polarised states, with the sign of the average choice determined by that of the population's IWA. If the system was already in a polarised state at zero IWA, while the sign and value  of the average choice will in general be unambiguously determined by that of the field, for small enough values of the latter, the symmetry is not completely broken and the degeneracy of the average choice equilibria persists.  While no longer equiprobable or symmetric, depending on the previous state of equilibrium of the system, the final average magnetisation may not be aligned with the IWA. Metastable states can be understood as a collectively reinforced choice that persists even when the rest of conditions are ripe for a change in the sign of the average choice. In economic demand contexts, it involves two coexisting low and high demand states for a given price. They are associated to a first order phase transition at $h=0$, which provides a mechanism for abrupt, discontinuous changes in the sign of the average choice and the possibility for \emph{hysteresis} (changes in the state of the system when varying the parameters that can not be undone by reversing the process).  

Some attempts at explaining data referring to real social processes using this type of approach have been made \cite{Glaeser1996, Weisbuch2000, Michard2005, Bernardes2002, Fortunato2007, Galam2003, Borghesi2007}. This line of research remains, however, somewhat scarce, and more efforts need to be put in this direction.

In this work, besides considering social influence, two different binary choice variables, each governed by Ising mean field dynamics, are coupled in another ferromagnetic like interaction. These and similar setups have been studied in physical contexts, particularly for their interest in explaining plastic phase transitions \cite{Imry1975, Galam1988, Korutcheva1988, Galam1989, Galam1990, Galam1995}. It amounts to the introduction of an additional contribution to the utility \eqref{eq:utility} that bonuses or penalises agents for the alignment of both spin variables. This is the case of interest when considering two different groups making the same choice when both groups are interdependent in their decision making process. Each individual is subject to social pressure to conform to its own group's average. Besides, agents will also tend to align ($k>0$) or dis-align ($k<0$) their options with the other group's average choice. This setup is referred to as the \emph{group interdependence} or the \emph{nonlocal model} (agents can be considered to interact with all members from its own group and the other one). Coupled spin variables are also the case of interest to study a single group in which each individual has to make two choices which are related, which can be labelled \emph{individual interdependence or local model} (the coupling between both spin variables is then only through each individual).  These systems can be in three possible phases depending on the two (coupled) order parameters: \emph{unpolarised} or \emph{paramagnetic phase} ($s=t=0$), \emph{polarised} or \emph{ferromagnetic} phases ($s\neq0$, $t\neq0$) and \emph{mixed phases} (one zero and one nonzero order parameter)\footnote{This terminology convention is different from that used in \cite{Galam1995}, where the ferromagnetic phase is referred to as mixed phase.}.

In the next two sections, Hamiltonians for both models are described and expressions for their free energies and systems of equations of state analytically derived. Section \ref{sec:phadia} describes and compares the resulting phase diagrams for the zero opinion fields or unbiased case. Finally, some concluding remarks are made.

\section{The group interdependence or nonlocal model}

Let us consider two groups, one made up of $N_{s}$ agents making binary choice $s_{i}=\pm 1$ and the other of $N_{t}$ agents making binary choice $t_{i}=\pm 1$ with dynamics governed by the Hamiltonian:

\begin{equation}
\label{eq:ham2p1cho1}
H = \sum_{(i,j)}\left(-\frac{J_{s}}{N_{s}}s_{i}s_{j}-\frac{J_{t}}{N_{t}}t_{i}t_{j}-\frac{k(N_{s}+N_{t})}{2N_{s}N_{t}}s_{i}t_{j}\right)-\sum_{i}\left(h_{s}s_{i}+h_{t}t_{i}\right)
\end{equation} 

\noindent where summations over $(i,j)$  are $1\leq i<j \leq N_{s}$ for the first term, $1\leq i<j \leq N_{t}$ for the second term and $1\leq i\leq N_{s}$, $1\leq j\leq N_{t}$  for the mixed term. Summations over $i$ are $1 \leq i \leq N_{s}$ for the fourth term and $1 \leq i \leq N_{t}$ for the last term.

The \emph{intra-couplings} $J_{s}$, $J_{t}$ are positive and constant and quantify the strength of social pressure within each group. Each individual's decision is also affected by the agent's correct perception of what the other group is doing through  the \emph{inter-coupling} $k$. It may either reinforce ($k>0$) or discourage ($k<0$) alignment of choices between individuals in different groups but interaction is always symmetric (exactly reciprocal).  Using mean field theory, which is exact for infinite groups (thermodynamic limit in physical terminology), is equivalent to considering each agent as influenced by its correct perception of the average choice of the other group. Besides intra- and inter- group social interactions, each group is subject to a constant opinion field. The most natural application seems to problems where the interaction between both groups is of social nature. It could also refer, however, to actual gains or losses for agents unrelated to social interaction.

This setup could be used to study problems such as the incidence of social pathologies in neighbouring districts, public opinion on a certain issue in two neighbouring countries, technology choice in two different related industries, female participation in the labour market in two neighbouring regions, etc.

When both groups are of the same size  (which is the case when considering infinitely large groups) this is the system studied in \cite{Korutcheva1988}. This work extends the results presented there using numerical methods to calculate stable solutions of the system in thermodynamic equilibrium.

For $N_{s}=N_{t}=N$, using mean field theory (on all intra- or inter-coupling terms) we can rewrite \eqref{eq:ham2p1cho1} as:

\begin{equation}
\label{eq:ham2p1cho2}
H = \frac{NJ_{s}}{2}s^{2}+\frac{NJ_{t}}{2}t^{2}+Nkst
-\left(J_{s}s+kt+h_{s}\right)\sum_{i}s_{i} -\left(J_{t}t+ks+h_{t}\right)\sum_{i}t_{i} 
\end{equation} 

\noindent where $s=\frac{1}{N}\sum_{i}s_{i}$ and $t=\frac{1}{N}\sum_{i}t_{i}$  are the average choices in each of the groups. The corresponding partition function for the representative canonical ensemble ($Z=Tr e^{-\beta H}$ where $Tr$ indicates sum over all possible spin configurations) can be expressed  

\begin{equation}
Z = e^{-\beta(\frac{N}{2}s^{2}J_{s}+\frac{N}{2}t^{2}J_{t}+Nkst)} 
\lbrack 2\cosh\left(\beta\left(J_{s}s+kt+h_{s}\right)\right)2\cosh\left(\beta\left(J_{t}t+ks+h_{t}\right)\right) \rbrack^{N}
\end{equation}

\noindent where $\beta$ is the social permeability or inverse of the temperature (which in this case accounts for statistical fluctuations).
 
Finally the system's free energy density ($f=F/N$ with $F$ the free energy $F= \beta^{-1}\log(Z)$) will be given in the mean field approximation by
\begin{equation}
f = \frac{1}{2}J_{s}s^{2}+\frac{1}{2}J_{t}t^{2}+kst-\frac{1}{\beta}\ln\left(2\cosh\left(\beta\left(J_{s}s+kt+h_{s}\right)\right)\right) 
-\frac{1}{\beta}\ln\left(2\cosh\left(\beta\left(J_{t}t+ks+h_{t}\right)\right)\right)
\label{eq:freeE}
\end{equation}
 
Stable states of the system will be those minimising the free energy. After derivation and simple algebraic manipulation, critical points can be found to be given  by  the system of equations of state:

\begin{equation}
\label{eq:eqstanlo}
\begin{array}{l}
a\left(s-\tanh\left(\beta\left(J_{s}s+kt+h_{s}\right)\right)\right) = 0  \\
a\left(t-\tanh\left(\beta\left(J_{t}t+ks+h_{t}\right)\right)\right) = 0
\end{array}
\end{equation}

\noindent with $a = J_{s}J_{t}-k^{2}$.

There are thus two different cases. For the \emph{degenerate case} ($a=0$), substituting $J_{t}=J_{s}/k^{2}$ in the original system both give the same equation of state:

\begin{equation}
J_{s}s+kt-J_{s}\tanh\left(\beta\left(J_{s}s+kt+h_{s}\right)\right) 
-k\tanh\left(\beta\left(\frac{k^{2}}{J_{s}}t+ks+h_{t}\right)\right) = 0 
\label{eq:eqstadeg}
\end{equation}

For the  \emph{non-degenerate case} ($a \neq 0$), on which this work focuses, critical points will be given by solutions to the system of equations of state:

\begin{equation}
\begin{array}{l}
s = \tanh\lbrack\beta\left(J_{s}s+kt+h_{s}\right)\rbrack\\
t = \tanh\lbrack\beta\left(J_{t}t+ks+h_{t}\right)\rbrack
\end{array}
\label{eq:eqstandeg}
\end{equation}

Positive definite Hessian (evaluated at the solution of the system \eqref{eq:eqstandeg}) is a sufficient condition for a critical point to be a minimum. In this case, its determinant can be written as: 

\begin{equation}
\label{eq:dethessnolo}
\det(\mathcal{H})=a\left(1-\beta J_{s}\gamma_{s}-\beta J_{t}\gamma_{t}\right) +\beta^{2}a^{2}\gamma_{s}\gamma_{t}
\end{equation}

\noindent and the free energy's second derivative (and first diagonal element of the Hessian) as:
\begin{equation}
\label{eq:fssnolo}
\frac{ \partial^{2}f}{\partial s^{2}} = J_{s}-\beta J_{s}\,^{2}\gamma_{s}-\beta k^{2}\gamma_{t} 
\end{equation}

\noindent where $\gamma_{s}=\frac{1}{\cosh^{2}\lbrack \beta \left(J_{s}s+kt+h_{s}\right)\rbrack}$ and $\gamma_{t}=\frac{1}{\cosh^{2}\lbrack \beta \left(J_{t}t+ks+h_{t}\right)\rbrack}$. Positive \eqref{eq:dethessnolo} and \eqref{eq:fssnolo} are sufficient conditions for minima.

When zero opinion fields are considered, polarised solutions of the system of equations of state \eqref{eq:eqstandeg} will always appear in pairs $(s,t)$ and $(-s,-t)$, each pair of solutions having the same stability. There is also a symmetry relating solutions under a change in sign in $k$ and in any one of the average magnetisations. Furthermore, the  unpolarised state or paramagnetic solution $s=t=0$ is always a critical point of $f$. Critical regions where its stability changes can be studied by linearising equations \eqref{eq:eqstandeg} for $|s| \ll 1$ and $|t| \ll 1$  (at finite non zero temperature) when $h_{s}=h_{t}=0$, yielding:

\begin{equation}
\begin{array}{l}
s = \beta\left(J_{s}s+kt\right) + O(s^{3},t^{3}, s^{2}t, st^{2})\\
t = \beta\left(J_{t}t+ks\right) + O(s^{3},t^{3}, s^{2}t, st^{2})
\end{array}
\end{equation}.

\noindent Further simplification of this system leads to the expression:
\begin{equation}
a\beta^{2}-(J_{s}+J_{t})\beta + 1 = 0
\label{eq:Tceq}
\end{equation}.

Solving \eqref{eq:Tceq} gives two values of $\beta$ where the stability of the paramagnetic phase changes from saddle point to maximum/minimum:

\begin{equation}
\label{eq:Tc}
\beta =\frac{J_{s}+J_{t} \pm \sqrt{(J_{s}+J_{t})^{2}-4a}}{2a}=\frac{J_{s}+J_{t} \pm \sqrt{(J_{s}-J_{t})^{2}+4k^{2}}}{2a}
\end{equation}

In the \emph{strong coupling regime ($k^{2}>J_{s}J_{t}$)}, there is a single physical (positive) value, where the  unpolarised solution changes from saddle point to maximum, and so of no relevance in the discussion of phase transitions. 

In the \emph{weak coupling regime ($k^{2}<J_{s}J_{t}$)}, there will be two physical values of  $\beta$. The larger one (given by the $+$ option in \eqref{eq:Tc}) is of the same type as discussed above. The smaller one (given by the $-$), is a critical point at which the unpolarised or paramagnetic phase will change from saddle to minimum and a second order phase transition takes place. For temperatures bellow the critical one, the system is in a polarised or ferromagnetic phase. This means that, even in the absence of personal predispositions of any type, social interaction both with individuals from within the agent's own group and from the other one,  promotes the emergence of a tendency to  polarisation (which is only complete,  i.e., $s=\pm 1$ and $t=\pm 1$ at $T=0$). The actual direction of choice is not determined, only the alignment or dis-alignment between both average choices (there are two physically equivalent states of equilibrium). Note that when opinion fields are {\it turned on}, even if the symmetry is broken and the paramagnetic solution will no longer be a critical point, this will still delimit regions of the phase diagram where interaction in itself has an impact on the aggregate outcome.

\section{The individual interdependence or local model}

Let us now consider a single group of $N$ identical individuals all of which are simultaneously making two interdependent binary choices $s_{i}=\pm 1$ and $t_{i}=\pm 1$. There will be an additional term in the utility function that penalises or bonuses the agent for aligning or not its two choices. Both of the decisions are subject to social influence to conform to the norm and to a constant opinion field. A Hamiltonian describing such a system is:

\begin{equation}
\label{eq:ham1p2cho1}
H = \sum_{(i, j)}\left(-\frac{J_{s}}{N}s_{i}s_{j}-\frac{J_{t}}{N}t_{i}t_{j}\right)-\sum_{i}\left(ks_{i}t_{i}+h_{s}s_{i}+h_{t}t_{i}\right)
\end{equation} 

\noindent where summations over $(i,j)$  are $1\leq i<j \leq N$ and summations over $i$ are $1 \leq i \leq N$. Note that in this case the interaction between both choice variables cannot be of social nature. 

Discrete choice problems of interest to the social sciences that can be explored using this model include the relation between teenage pregnancy and school dropout in a given population, results in simultaneous referendums or in elections in a bipartidist system, demand for two different software products of the same brand, motherhood and female participation in the labour market and so on.

This is  a similar model to that studied in \cite{Galam1995}, where random inter-coupling $k$ is considered. Most results presented, however, are for symmetric probability distributions $p(k_{i})$ which seem of little interest in the socioeconomic context, as they involve populations for which half of its members have a positive interaction between both choices while for the other half they tend to oppose. 

Hamiltonian \eqref{eq:ham1p2cho1} can be rewritten:
\begin{equation}
\label{eq:ham1p2cho2}
H = \frac{J_{s}+J_{t}}{2}-\frac{J_{s}}{2N}\left(\sum_{i}s_{i}\right)^{2}-\frac{J_{t}}{2N}\left(\sum_{i}t_{i}\right)^{2}-\sum_{i}\left(ks_{i}t_{i}+h_{s}s_{i}+h_{t}t_{i}\right)
\end{equation}

Following \cite{Galam1995}, the partition function of the representative canonical ensemble can be computed exactly for an  infinitely large system (thermodynamic limit) yielding:

\begin{equation}
Z=\frac{\beta N}{2\pi}(J_{s}J_{t})^{\frac{1}{2}}e^{-\frac{\beta}{2}(J_{s}+J_{t})}\int_{-\infty}^{\infty}\int_{-\infty}^{\infty}\,ds\,dt e^{-\beta Ng(s,t)}
\end{equation}

\noindent where $g(s,t)$ is the free energy functional such that the free energy density $f= \lim_{N\to\infty}\frac{F}{N}=\int_{-\infty}^{\infty}\int_{-\infty}^{\infty}\,ds\,dt\, g(s,t)$ and is given by the expression:

\begin{equation}
\label{eq:g}
g = \frac{1}{2}J_{s}s^{2}+\frac{1}{2}J_{t}t^{2} -\frac{1}{\beta}\ln\lbrack 2e^{\beta k}\cosh\left(\beta\left(J_{s}s+J_{t}t+h_{s}+h_{t}\right)\right)
+2e^{-\beta k}\cosh\left(\beta\left(J_{s}s-J_{t}t+h_{s}-h_{t}\right)\right)\rbrack
\end{equation}

Critical points of the free energy $f$ are those of the functional $g$. Introducing the notation $\alpha_{s}=\tanh\left(\beta(J_{s}s+h_{s})\right)$, $\alpha_{t}=\tanh\left(\beta(J_{t}t+h_{t})\right)$ and $\alpha_{k}=\tanh\left(\beta k\right)$, the system's equations of state can be written as:

\begin{equation}
\label{eq:1p2choeqstasim}
\begin{array}{l}
s = \displaystyle{\frac{\alpha_{s}+\alpha_{t}\alpha_{k}}{1+\alpha_{s}\alpha_{t}\alpha_{k}} }\\
t = \displaystyle{\frac{\alpha_{t}+\alpha_{s}\alpha_{k}}{1+\alpha_{s}\alpha_{t}\alpha_{k}} }
\end{array}
\end{equation}

In this case the Hessian's determinant takes the form:

\begin{equation}
\label{eq:locdethess}
det(\mathcal{H})=J_{s}J_{t}\lbrack 1- \frac{\beta \left( J_{t}\gamma_{t}(1-\alpha_{s}^{2}\alpha_{k}^{2})+J_{s}\gamma_{s}(1-\alpha_{t}^{2}\alpha_{k}^{2})\right)}{(1+\alpha_{s}\alpha_{t}\alpha_{k})^{2}}+
\frac{\beta^{2}J_{s}J_{t}\gamma_{s}\gamma_{t}\left((1-\alpha_{s}^{2}\alpha_{k}^{2})(1-\alpha_{t}^{2}\alpha_{k}^{2})-\gamma_{s}\gamma_{t}\alpha_{k}^{2}\right)}{(1+\alpha_{s}\alpha_{t}\alpha_{k})^{4}}\rbrack
\end{equation}

\noindent and $g$'s second derivative with respect to $s$ is given by:
\begin{equation}
\frac{ \partial^{2} g}{ \partial s^{2}}  = J_{s}-\frac{\beta J_{s}^{2}\gamma_{s}\left(1-\alpha_{t}^{2}\alpha_{k}^{2}\right)}{(1+\alpha_{s}\alpha_{t}\alpha_{k})^{2}} \label{eq:locgss}
\end{equation}

The zero opinion field case has the same symmetries described for the nonlocal non-degenerate zero field model. The paramagnetic or unpolarised solution is also always a critical point of $g$. Linearization of \eqref{eq:1p2choeqstasim} for $|s| \ll 1$ and $|t| \ll 1$ (at finite nonzero temperature) when $h_{s}=h_{t}=0$ yields:

\begin{equation}
\begin{array}{l}
s = \beta J_{s}s+\beta J_{t}\tanh(\beta k)t + O(s^{3},t^{3}, s^{2}t, st^{2})\\
t = \beta J_{t}t+\beta J_{s}\tanh(\beta k)s + O(s^{3},t^{3}, s^{2}t, st^{2})
\end{array}
\end{equation}

\noindent which can be simplified to

\begin{equation}
\label{eq:locTc}
J_{s}J_{t}\left(1-\tanh^{2}(\beta k)\right)\beta^{2}-(J_{t}+J_{s})\beta+1 = 0
\end{equation}

Depending on the values for the intra- and inter-couplings, there can be three qualitatively different scenarios when analysing equation \eqref{eq:locTc}, with either one, two or three different solutions in $\beta$ (all positive and thus physically relevant). In all cases, only the smaller (or only) one of these (the highest temperature) will be a critical point where the paramagnetic or unpolarised phase becomes unstable and there is a second order phase transition taking place.

\section{Phase diagrams for the unbiased case}
\label{sec:phadia}

The Newton$-$Raphson algorithm has been used to numerically solve the equations of  state for the unbiased case ($h_{s}=h_{t}=0$), fixed $J_{s}=1$ in some given units (which is equivalent to measuring the rest of couplings and fields in terms of $J_{s}$) and different values of the rest of the parameters. Python libraries developed for the computation and analysis of these are available at {\tt https://github.com/anafrio/}. Results are summarised through figures of the phase diagram's cross-sections for both models for specific values of the rest of parameters. These have been chosen for the sections to be representative of all the possible phenomenology. More details on the solutions found (including unstable ones) and dependence on the different parameters can be found in \cite{Rio2010}. 

Unbiased situations refer to what has been loosely referred to as trends or traditions, i.e., social conventions which will give no particular advantage or disadvantage to the isolated individual and choice. For the group interdependence model, this would be the case, for example, to study the use of a particular outfit or accessory in two different socioeconomic groups, or the adoption of specific vocabulary (argot, technical term) by two such groups. Unbiased individual interdependence could be of use to tackle problems such as how the use of an accessory in a group will be affected by the use of another when it is considered to be {\it fashionable or unfashionable} to wear them together.

The study of the unbiased case is also useful to get some insight on what will happen in a more general situation of biased homogeneous groups. The introduction of nonzero constant opinion fields, will in general determine the sign of the average magnetisation and break the equilibria degeneracy, and the unpolarised state will no longer be stable. What were unpolarised regions for the unbiased case, will become characterised by an equilibrium demand or acceptance univocally determined by the constant IWA (or IWB and price). In the already polarised regions before the introduction of fields, however, multiple possible equilibria and first order phase transitions can be expected at low enough IWAs and temperatures \cite{Rio2011}.

\begin{figure}
\centering
\subfloat[]{\includegraphics[width=0.5\textwidth]{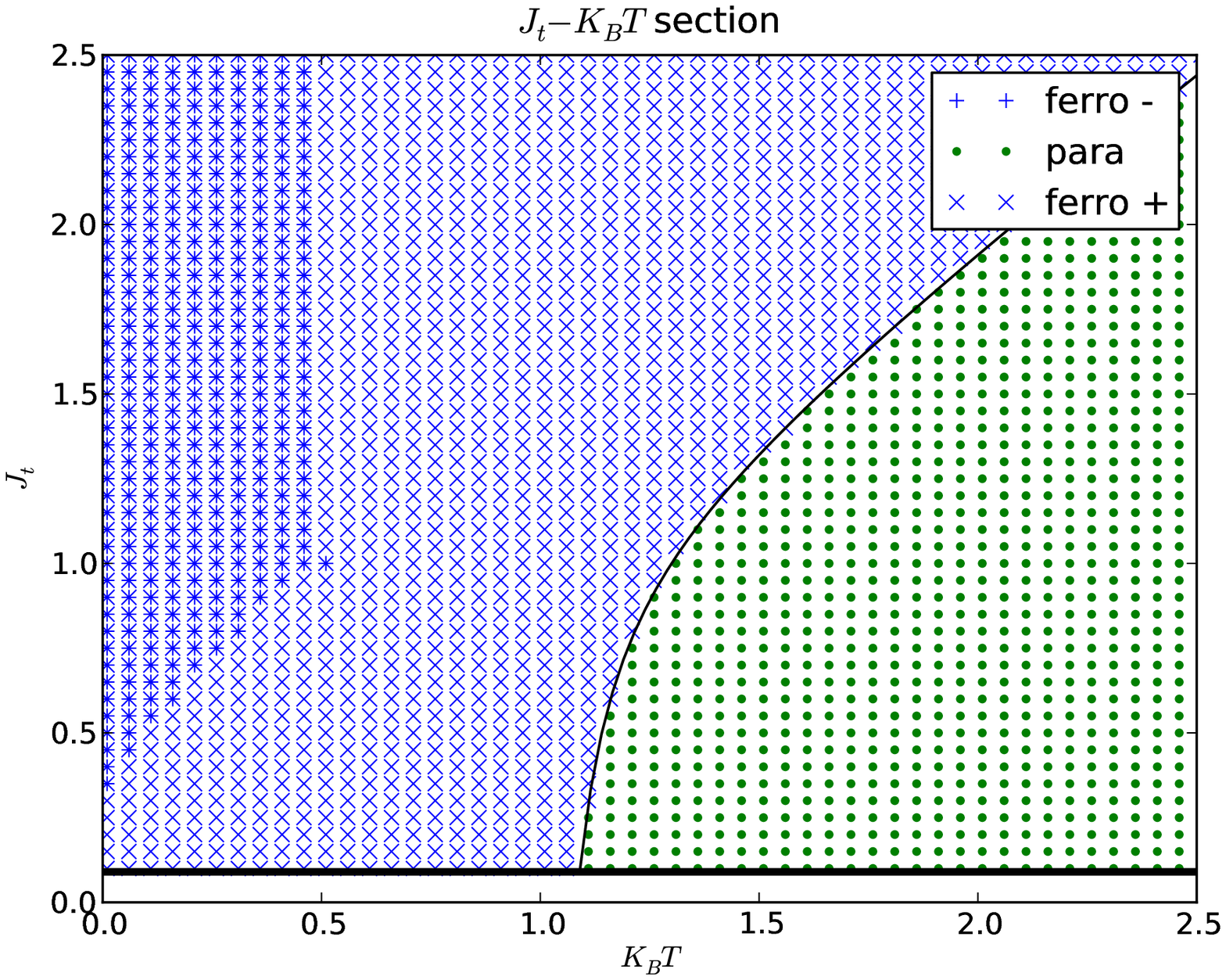}}
\subfloat[]{\includegraphics[width=0.5\textwidth]{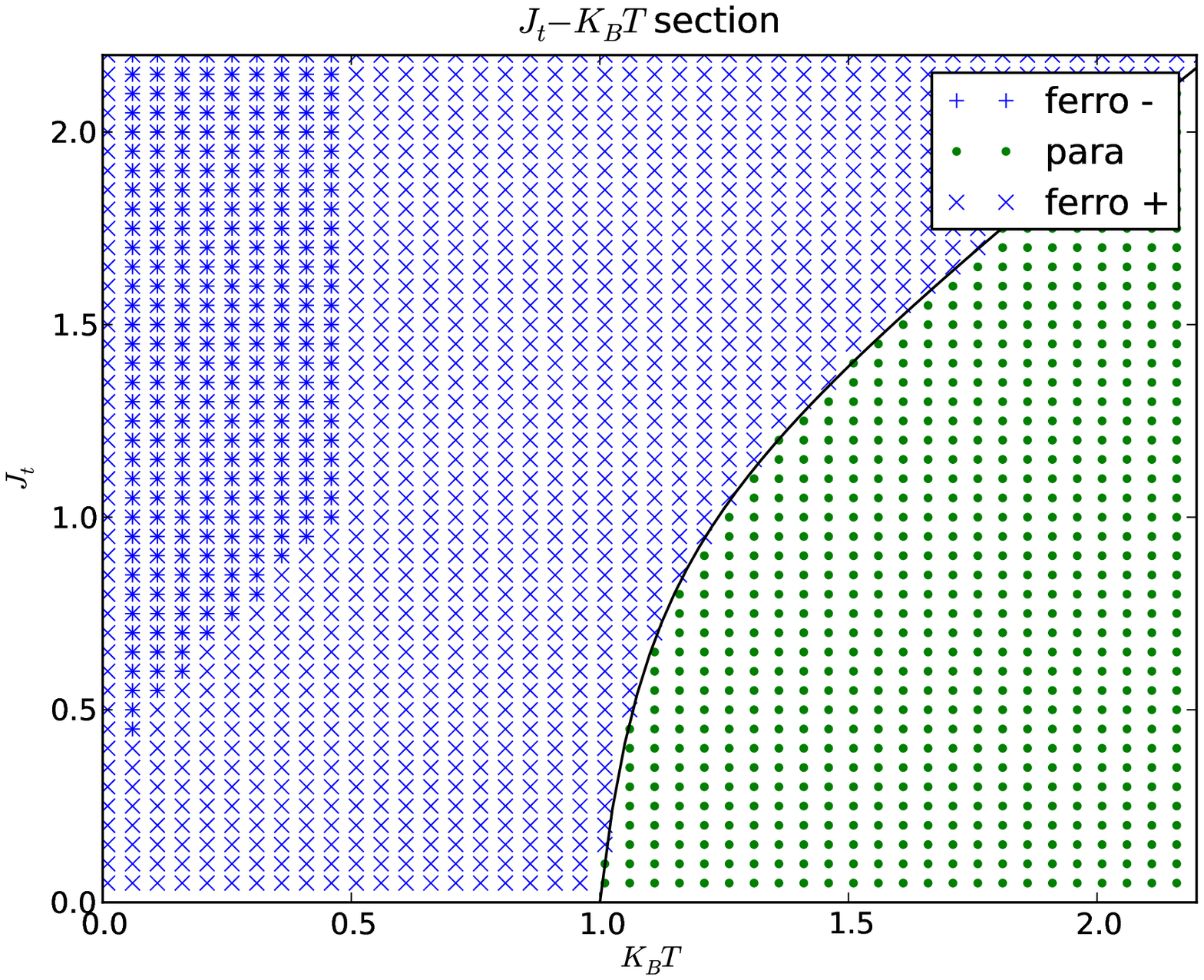}}
\caption{$J_{t}- \beta^{-1}$ section for the (a) nonlocal and (b) local models for fixed $J_{s}=1$ and $k=0.3$. }
\label{fig:phadiaJT}
\end{figure}

\begin{figure}
\centering
\subfloat[]{\includegraphics[width=0.5\textwidth]{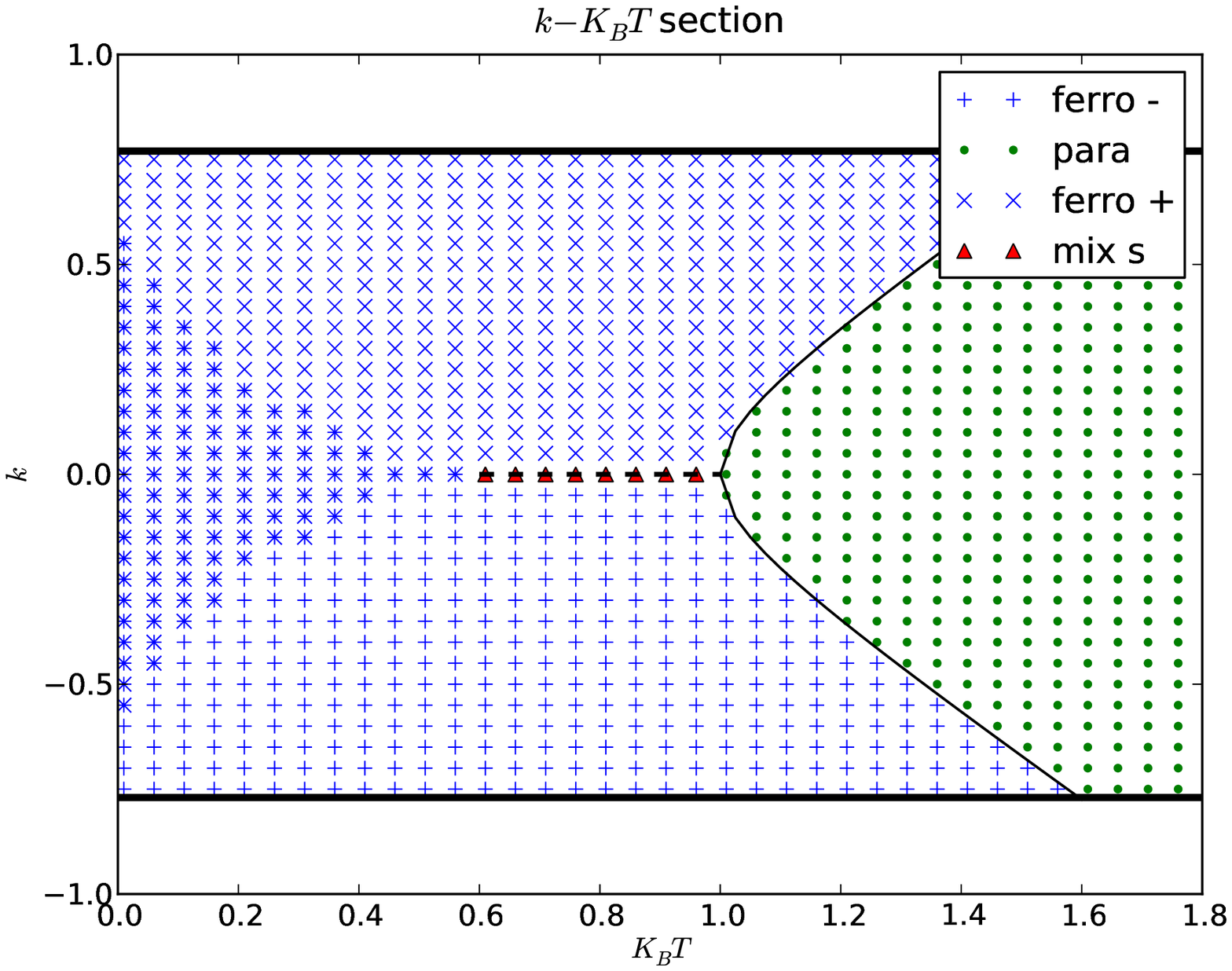}}
\subfloat[]{\includegraphics[width=0.5\textwidth]{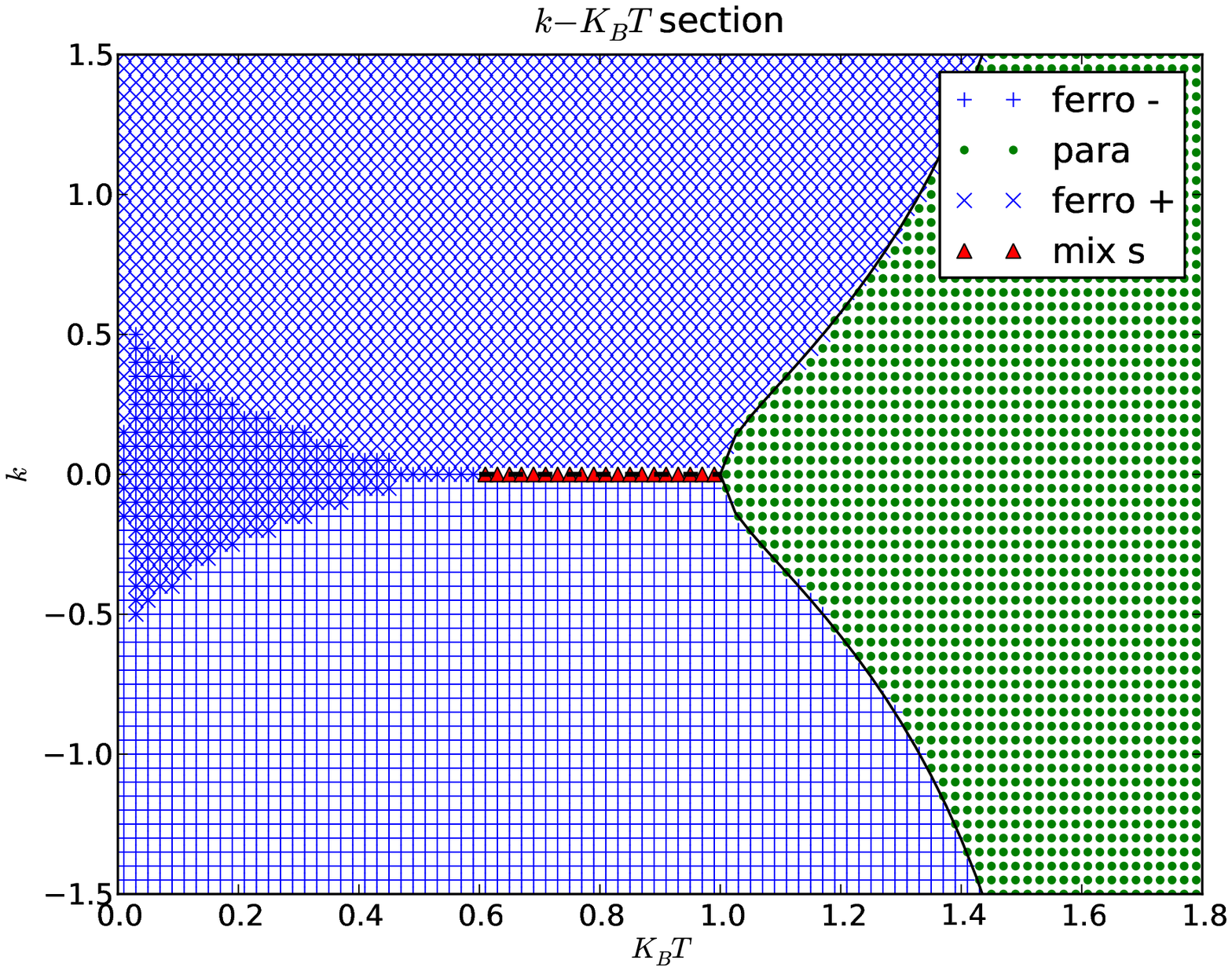}}
\caption{$k-\beta^{-1}$ section for the (a) nonlocal and (b) local models for fixed $J_{s}=1$ and $J_{t}=0.6$.}
\label{fig:phadiakT}
\end{figure}

\begin{figure}
\centering
\subfloat[]{\includegraphics[width=0.5\textwidth]{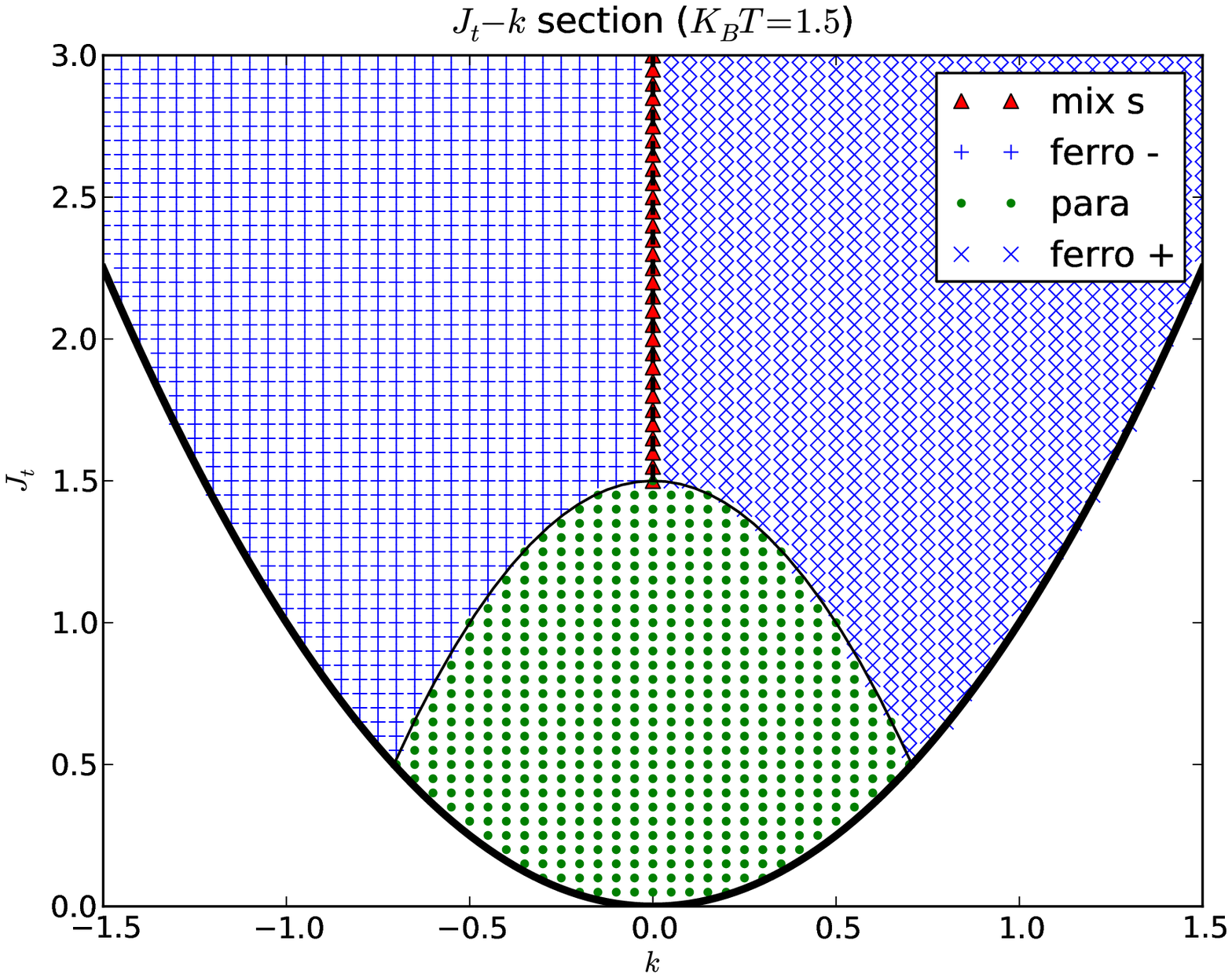}}
\subfloat[]{\includegraphics[width=0.5\textwidth]{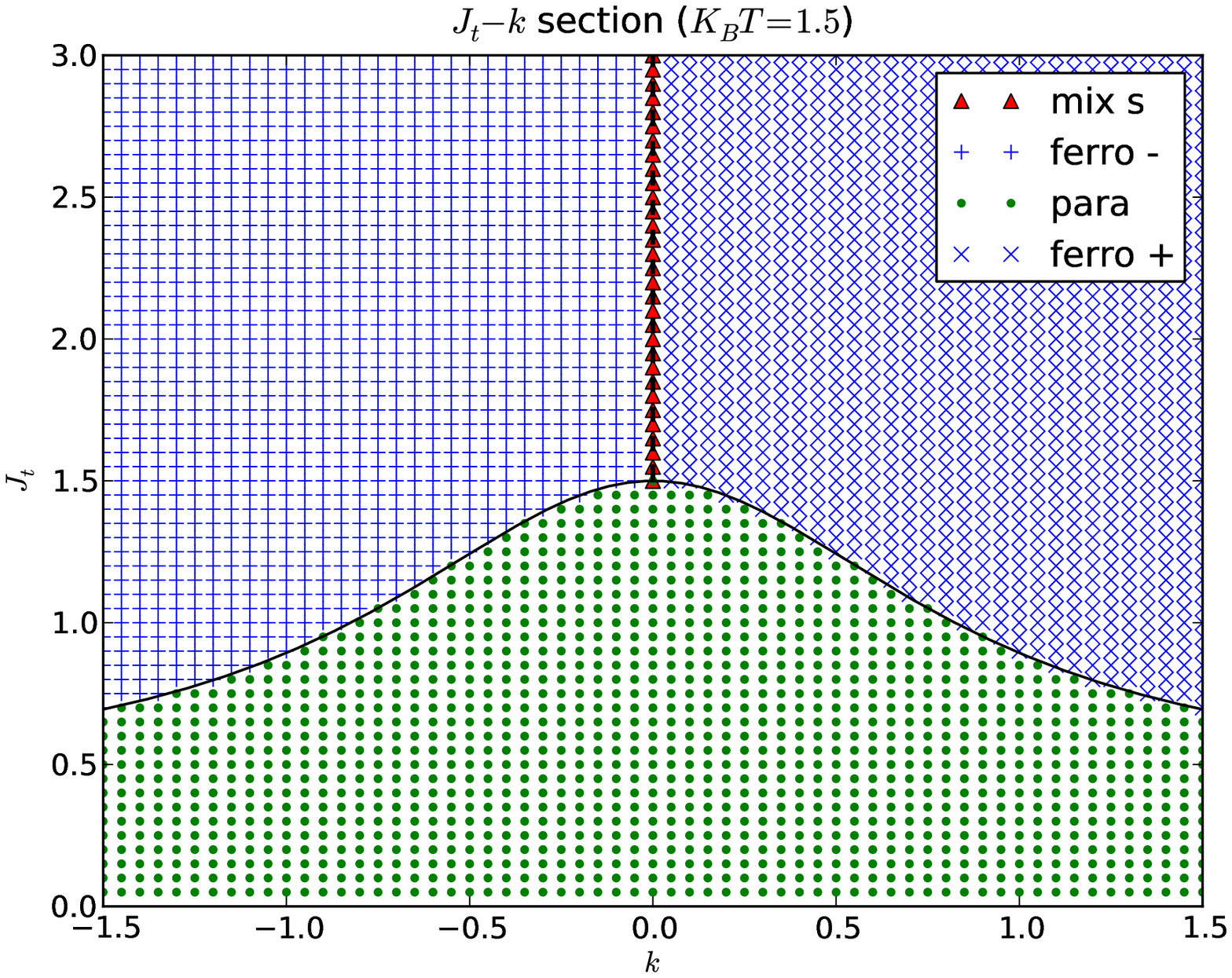}}\\
\subfloat[]{\includegraphics[width=0.5\textwidth]{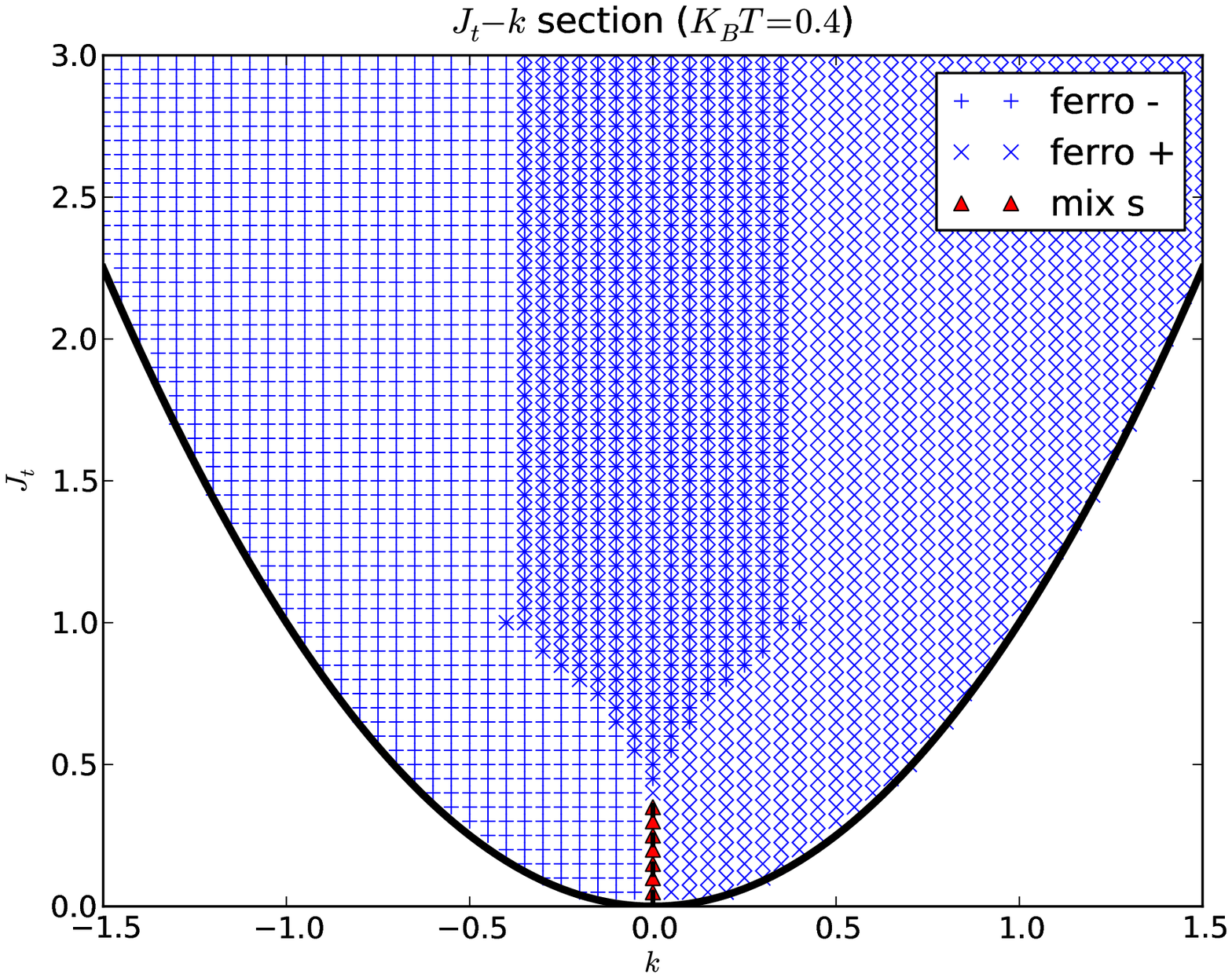}}
\subfloat[]{\includegraphics[width=0.5\textwidth]{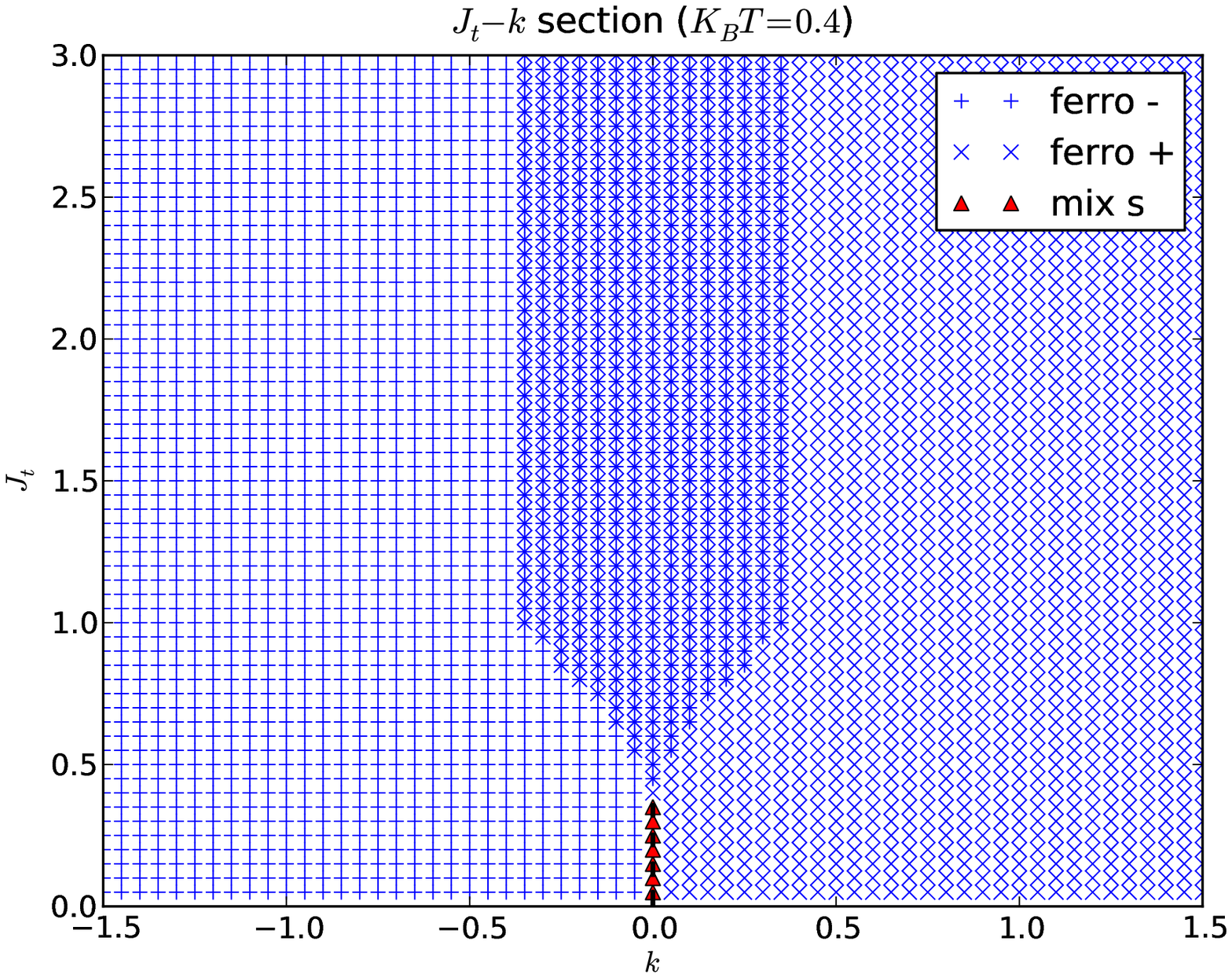}}
\caption{ $J_{t}-k$ sections for the (a) nonlocal and (b) local models at $\beta^{-1}=1.5$ and for the (c) nonlocal and (d) local models at $\beta^{-1}=0.4$. In all cases $J_{s}=1$.}
\label{fig:phadiaJk}
\end{figure}

Figures \ref{fig:phadiaJT} to \ref{fig:phadiaJk} show numerically calculated solutions which are stable (minima of the free energy). Green dots are used for the paramagnetic or unpolarised phase (only one stable equilibria), blue asps (x) for ferromagnetic or polarised phases with both average choices of the same sign and crosses (+) for ferromagnetic  or polarised phases with average choices of opposed signs (two physically equivalent stable equilibria in each case). Red triangles show mixed phase solutions (also two equilibria). When both types of ferromagnetic or polarised phases are superposed, there are four different equilibria, two stable (those aligned according to the sign of $k$) and two metastable.  

The \emph{mixed segment} is shown with a dashed black line (figures \ref{fig:phadiakT} and  \ref{fig:phadiaJk}). This is the only region where mixed phases are solutions and is the same for both models as these can only exist for $k=0$ (uncoupled case) and when one of the Ising models is in the paramagnetic phase while the other is still in the ferromagnetic phase, i.e., $J_{t}< \beta^{-1}<J_{s}$ when $J_{t}<J_{s}$ (or $J_{s}< \beta^{-1}<J_{t}$ when $J_{s}<J_{t}$). Whenever there is a nonzero value of $k$, it is no longer possible for  the opinion to remain unpolarised if the other group/opinion to which it is coupled is not.

For the group interdependence or nonlocal model, the \emph{degenerate curve} $k^{2}=J_{s}J_{t}$ is drawn as a solid thick black line (figures \ref{fig:phadiaJT} (a), \ref{fig:phadiakT} (a) and \ref{fig:phadiaJk} (a) and (c)). As shown in the figures, there are no stable solutions of any type in the strong coupling regime. Frustration prevents the system from arriving at equilibrium. This was already being indicated by the fact that neither the completely polarised ($s=\pm1,t=\pm1$) solutions at zero temperature or the unpolarised one at very large temperatures are stable (which can be checked analytically).  When considering that the interaction between both groups is of pure social type, assuming stronger social pressure from within the group than from the outside seems pretty natural. It can be considered a nice feature of the model that it {\it breaks down} when the problem is, sociologically speaking, ill defined (groups have not been chosen appropriately). The coupling between both groups, however, needs not to be of social nature. In this case, this would be pointing at a real impossibility of arriving at a stable equilibria if the interdependence between both groups is the main driver of the choice making process. 

The \emph{critical curve}, where a second order phase transition takes place, is drawn in thin solid black (figures \ref{fig:phadiaJT}, \ref{fig:phadiakT} and \ref{fig:phadiaJk} (a) and (b)). This is given by the condition $\beta^{-1}>\max\{J_{s},J_{t}\}$ together with  \eqref{eq:Tc} for the nonlocal model or \eqref{eq:locTc} for the local model. It separates regions of polarised opinion (ferromagnetic phases) from others where the paramagnetic phase is the only stable equilibria (half of the population deciding each way for both choices for sufficiently large statistical fluctuations). When nonzero, constant IWAs are introduced, this will still separate regions where there will be a single possible value for the average choice from those where metastable equilibria can be expected to appear for low fields. Under such circumstances, drastic and irreversible shifts in the average opinion can be expected \cite{Rio2011}.

Figure \ref{fig:phadiaJT} shows a $J_{t}- \beta^{-1}$ section for the nonlocal (a) and local (b) models for $J_{s}=1$ and $k=0.3$. They are remarkably similar qualitatively and quantitatively. If both groups/choices were not coupled, the critical curve would be the line of slope one across the origin and in all of the ferromagnetic region there would be two possible values (of same absolute value) for each average magnetisation (four possible states when combined), all with the same probability. The introduction of interdependence between both groups/choices promotes polarisation, making larger statistical fluctuations (temperature) needed as compared to the uncoupled case for imitation to promote a certain collective trend. For high enough social permeabilities $\beta$, there will no longer be a threshold value of the social influence bellow which the system is in a paramagnetic phase. In the individual dependence case, even when we consider no imitation whatsoever for one of the choices ($J_{t}=0$), social pressure on the other is enough for polarisation to appear. This is not the case of the group dependence model, where zero intra-couplings are {\it forbidden}, as they always lay in the strong coupling regime.

When considering ferromagnetic regions, the introduction of the coupling will, in general, fix the relative signs between both choices, although not the direction. For example, if we consider the adoption of a certain technical term by two professional groups or academic disciplines, if these are close and there is affinity between them, the new term will either gain acceptance or not in both groups. Similarly, if both groups perceive each other as confronted in any way, a high acceptance in one of them will always come with a low acceptance in the other. Likewise, if we analyse teenage outfit, if two accessories are perceived as {\it fitting well}, they will either both be used by over half of the population or by less than half of it, with no possibility of one of the trends being successful without the other.

 For low enough temperatures and/or strong enough social pressure (coupling $J_{t}$), there is a region of metastability and the possibility of hysteresis. This behaviour is related to a first order phase transition at $k=0$. If the inter-coupling is slowly reversing its sign, the system may not immediately change in the alignment of average choices, i.e., the system may get trapped in a local minima of the free energy ({\it for a while}, statistical fluctuations will make the system {\it decay} to the real ground state eventually). This is never the case for uncoupled groups/choices in the unbiased case and can be given interesting interpretations when considering two groups/choices with mutual influence slowly changing from negative to positive (or positive to negative). In the example of vocabulary adoption, this will imply the possibility of the use of the word remaining low in one of the professional sectors even as the mutual perception becomes positive, before there is an abrupt change and becomes widely accepted or not in both groups. In this region, the introduction of weak nonzero fields will probably enrich the metastability landscape  \cite{Rio2011}.

Note that for the nonlocal model the metastability region extends up to $\beta^{-1}=0$ but it does not in the local case (which can be easily checked analytically). This difference can also be explained naturally in the light of social applications. Zero temperature is equivalent to a completely deterministic picture. Then, in the local case, each rational, utility maximiser knows exactly whether it is better off aligning or not both of its choices (although not in which direction). In the nonlocal case, however, although there is still a globally preferred condition of alignment or not between both groups, the choice is not up to every agent as the coupling depends on what the other group is doing.  

Figure \ref{fig:phadiakT} shows $k-\beta^{-1}$ sections for both models when $J_{s}=1$ and $J_{t}=0.6$. The uncoupled case is described by the $k=0$ axis. Besides the promotion of polarisation and the differences already noted when examining the $J_{t}- \beta^{-1}$ section, for the local model, the critical curve is such that $\lim_{|k|\to\infty} \beta^{-1}=J_{s}+J_{t}=1.6$. No similar asymptotic behaviour is shown by the nonlocal critical curve (for which high values of $|k|$ are of little interest as there are no stable solutions in the strong coupling regime). When the interrelation between both decisions is much more important than any of the imitation strengths, the critical value of social permeability at which a different qualitative picture can emerge due to social interaction, is basically determined by the intra-couplings. Further increasing $|k|$ will have little impact in the critical value of the temperature and, once permeability is fixed and while in a high $|k|$ range, small variations in the interdependence of both choices will never bring about a qualitative change in the average outcome.

For both models, if $\beta^{-1}>J_{s}+J_{t}$, the only stable state, if any, is the unpolarised one. In the nonlocal or group interdependence model, this is due to the intersection of the degenerate curve and the critical curve, precisely at $J_{s}+J_{t}$.  In the local or individual interdependence model, it has to do with the asymptotic behaviour of the critical curve described above. This also makes sense in our context: at low enough permeabilities, in the local model, the interdependence between both choices can be made as large as desired without making social influence relevant qualitatively. This means that once we consider constant, nonzero IWAs, if the statistical fluctuations considered are large enough (such that $\beta^{-1}>J_{s}+J_{t}$), both average choices will be aligned with the fields, no matter what the value of $k$ is, and no multiple equilibria or abrupt changes will be possible.

As for $J_{t}-k$ sections, the situation is qualitatively different depending on the temperature chosen. Figures \ref{fig:phadiaJk} (a) and (b) show the sections for the nonlocal and local models when $\beta^{-1}=1.5>J_{s}=1$. In this case we have mixed segments for $k=0$ and $J_{t}> \beta^{-1}=1.5$, paramagnetic phase (unpolarised collective opinions) bellow the critical curve and no metastability region. As before, the uncoupled case is described by the $k=0$ axis. Again, only the local case presents asymptotic behaviour of its critical curve $\lim_{|k|\to\infty}J_{t}\to \beta^{-1}-J_{s}=0.4$.   As there, for $J_{t}<\beta^{-1}-J_{s}$, only the unpolarised state will be stable (if any) for both models. If the dependence between both choices $k$ is strong enough, the critical value of $J_{t}$ will be practically independent of its value.

Figures \ref{fig:phadiaJk} (c) and (d) show the sections for the nonlocal and local models when $\beta^{-1}=0.4<J_{s}=1$. Mixed segments lie, for $k=0$ (uncoupled case), on $0<J_{t}<\beta^{-1}=0.4$. There is no paramagnetic region (both opinions will always be polarised). There is a region of metastability present for low enough $|k|$ and/or high enough $J_{t}$, with the same socioeconomic implications described above.

\section{Conclusions}

Two different coupled Ising models have been analysed in the mean field regime in the context of binary choices and socioeconomic interactions. One of them is naturally suited to study how two groups affect each other when both are making the same choice (group interdependence or nonlocal model). The other to the case of a group where individuals make two choices that are interrelated (individual interdependence or local model). Phase diagrams have been discussed for the zero choice fields case (unbiased populations). These can be useful when analysing the success of certain trends or traditions and is enough to delimit regions of the phase diagram where social and choice interaction will be important to the qualitative outcome when the fields are turned on.

When compared to the uncoupled unbiased case, the introduction of interdependence, to an exogenous average choice or to another decision the individual is making, shifts the system towards polarisation, that is, larger statistical fluctuations (or smaller social permeability) is required for the average choices to remain unaffected by the social and choice interaction. Either both groups/choices will be simultaneously polarised or unpolarised. In the former case, there will in general be, as in the uncoupled case, two possible values of opposite sign for each average choice. The difference is that now the relative sign between them is in general fixed by the introduction of the coupling. This bonds the success of any fashion or trend to the success (positive interdependence) or failure (negative interdependence) in the other group/fashion to which it is coupled. There is a region at low inter-coupling and fluctuations (physical temperature) where it is possible for both average choices to align differently from the coupling between them. This allows for sudden, irreversible shifts from low to high (or high to low) demand or acceptance (which was never the case in the uncoupled unbiased case), and for strong history dependence.

When comparing the effects of coupling the choice to the same decision making process in an external group to that of another choice each individual is making, there are also some interesting differences. The coupling of groups will prevent them from arriving at a state of stable equilibrium whenever it is the main driver of the decision making process. In the completely deterministic picture, only group interdependence will allow for the possibility of alignment between the average choices contradicting the sign of the coupling. Although, contrary to the case of group interdependence, there are stable states for high individual choice interdependence, at large enough values of the inter-coupling, the polarisation threshold will be independent of it and determined by the sum of social imitation on each choice.

The models presented are a first look into the problem and  share the limitations exposed for the uncoupled case in the introduction. Probably the greatest weakness of this approach is that of considering completely homogeneous populations. The first natural way to encode a group's particularities and preferences seems to be the use of random choice fields, which is part of work in progress \cite{Rio2011}. From a wider perspective, the most important line of work might be testing the suitability of these and other related models to explain -ideally quantitatively but at least qualitatively- data for real social processes. This is no simple task and there is currently also work being done on these lines \cite{Rio2011a}.

\section*{Acknowledgements}

The authors wish to thank Prof. N. Tonchev for discussions on the mean field analysis of both models and for indicating previous work of Galam, Salinas and Shapir on the local model. They are grateful to Prof. S. Galam for some interesting comments and suggestions for further reading which proofed to be enlightening. They also want to thank Daniel Guinea for his sociological insights and for the chance to work on real data.  This work has been financially supported by the Ministerio de Ciencia e Innovaci\'on (Spain), Project No. FIS2009-09870.

\newpage
\bibliographystyle{unsrt}
\bibliography{draft}

\end{document}